\documentclass[10pt,twocolumn]{article}

\usepackage[a4paper,margin=1in]{geometry}
\usepackage{times}
\usepackage{authblk}
\usepackage{graphicx}
\usepackage{amsmath,amssymb}
\usepackage{hyperref}
\usepackage{braket}
\usepackage{caption}
\usepackage{booktabs}
\usepackage{tikz}
\usepackage{subcaption}
\usepackage{float}
\usetikzlibrary{shapes.geometric, arrows.meta, positioning, fit, backgrounds}

\title{\bfseries Geometric Latent Space Tomography with Metric-Preserving Autoencoders}

\author[1]{S.M. Yousuf Iqbal Tomal}
\author[1]{Abdullah Al Shafin}

\affil[1]{Department of Computer Science and Engineering, BRAC University, Dhaka, Bangladesh}

\date{}

\begin{document}
\maketitle

\begin{abstract}

Quantum state tomography faces exponential scaling with system size, while recent neural network approaches achieve polynomial scaling at the cost of losing the geometric structure of quantum state space. We introduce geometric latent space tomography, combining classical neural encoders with parameterized quantum circuit decoders trained via a metric-preservation loss that enforces proportionality between latent Euclidean distances and quantum Bures geodesics. On two-qubit mixed states with purity 0.85--0.95 representing NISQ-era decoherence, we achieve high-fidelity reconstruction (mean fidelity $F = 0.942 \pm 0.03$) with an interpretable 20-dimensional latent structure. Critically, latent geodesics exhibit strong linear correlation with Bures distances (Pearson $r = 0.88$, $R^2 = 0.78$), preserving 78\% of quantum metric structure. Geometric analysis reveals intrinsic manifold dimension 6.35 versus 20 ambient dimensions and measurable local curvature ($\kappa = 0.011 \pm 0.006$), confirming non-trivial Riemannian geometry with $O(d^2)$ computational advantage over $O(4^n)$ density matrix operations. Unlike prior neural tomography, our geometry-aware latent space enables direct state discrimination, fidelity estimation from Euclidean distances, and interpretable error manifolds for quantum error mitigation without repeated full tomography, providing critical capabilities for NISQ devices with limited coherence times.
\end{abstract}
\section{Introduction}
Quantum state tomography (reconstructing density matrices from measurement data) is essential to verify quantum computers, characterize quantum channels, and validate quantum protocols. However, traditional tomography scales exponentially: full reconstruction of an $n$-qubit state requires $\mathcal{O}(4^n)$ measurements and computational resources, rendering the approach intractable beyond $\sim$10 qubits\cite{nielsen2000quantum,paris2004quantum}. This exponential barrier has driven decades of research into compressed sensing and adaptive methods, yet even these approaches struggle with measurement efficiency and computational cost in the many-qubit regime\cite{gross2010quantum,cramer2010efficient}.

Machine learning has recently opened a new paradigm for quantum state reconstruction. Neural network-based tomography achieves polynomial scaling by learning compressed state representations from training data, demonstrating impressive reconstruction fidelities while bypassing the exponential cost of maximum likelihood estimation. However, these methods treat quantum states as abstract vectors, optimizing solely for reconstruction accuracy while ignoring the geometric structure of the quantum state space\cite{torlai2017neural,carrasquilla2019reconstructing,lohani2020machine}. Quantum states naturally form a Riemannian manifold where the Bures metric, derived from the Uhlmann-Jozsa fidelity, defines the geodesic distance between density operators\cite{uhlmann1976transition,jozsa1994fidelity}. This geometry governs quantum distinguishability, encodes statistical properties through quantum Fisher information, and defines optimal state transformations\cite{braunstein1994statistical,petz1996monotone}. By ignoring this structure, current neural approaches sacrifice interpretability and lose access to geometric operations such as direct fidelity estimation and geodesic interpolation.

We introduce \textit{geometric latent space tomography}, which achieves both compression and geometric fidelity by learning neural representations that explicitly preserve the Bures metric structure. Our hybrid architecture combines a classical neural encoder with a parameterized quantum circuit decoder, trained using a novel \textit{metric-preservation loss} that enforces proportionality between Euclidean distances in latent space and Bures geodesics in quantum state space. This design embodies a key principle: if latent representations faithfully capture quantum geometry, then simple Euclidean operations (computing distances, finding nearest neighbors, interpolating along lines) correspond to meaningful quantum operations without requiring density matrix reconstruction. This capability is particularly valuable for NISQ devices where repeated full tomography is impractical due to limited coherence times and gate fidelities\cite{preskill2018}. The result is an interpretable latent manifold where proximity reflects quantum similarity, enabling geometry-aware analysis at computational cost orders of magnitude below traditional methods.

Our approach yields practical advantages: fast fidelity estimation from latent distances, quantum error mitigation through interpretable error manifolds where decoherence trajectories become visible geometric structures, state discrimination via simple nearest-neighbor queries, and geometric quantum machine learning where states are manipulated as points on a learned Riemannian manifold.

This work makes three key contributions. First, we introduce a metric-preserving hybrid architecture combining neural encoding with quantum circuit decoding and a novel training loss that explicitly enforces geometric fidelity. Second, we develop comprehensive validation tools including geodesic-Bures correlation analysis, intrinsic dimensionality estimation, and curvature measurement, establishing quantitative metrics for geometric fidelity beyond reconstruction precision. Third, we demonstrate that compression and metric preservation are compatible objectives, achieving 78\% geometric structure preservation while maintaining 94.2\% tomographic fidelity, with computational advantage scaling as $\mathcal{O}(d^2)$ versus $\mathcal{O}(4^n)$ for density matrix operations. Our results suggest a principled path toward interpretable quantum machine learning on geometric state representations.

\section{Methods}

\subsection{Quantum State Preparation and Measurement}

\subsubsection{Mixed State Generation}

We generate a diverse ensemble of 2-qubit mixed states with controlled purity in the range $\mathrm{Tr}(\rho^2) \in [0.85, 0.95]$, representing the practical regime for near-term quantum devices where decoherence is moderate. The training set comprises $N_{\text{train}} = 2000$ states, and the validation set contains $N_{\text{val}} = 500$ states.

To ensure diversity in the quantum state manifold, we employ seven distinct state preparation channels with equal sampling probability. Depolarized states are generated via $\rho_{\text{depol}}(p) = (1-p)|\psi\rangle\langle\psi| + \frac{p}{d}\mathbb{I}_d$, where $|\psi\rangle$ is sampled from the Haar measure on pure states, $d = 2^n$ is the Hilbert space dimension, and the depolarization parameter $p$ is determined by inverting the purity relation $\mathrm{Tr}(\rho_{\text{depol}}^2) = (1-p)^2 + \frac{p^2}{d} + \frac{2p(1-p)}{d}$. Werner states follow the form $\rho_{\text{Werner}}(p) = p|\Phi^+\rangle\langle\Phi^+| + \frac{1-p}{d}\mathbb{I}_d$, where $|\Phi^+\rangle = \frac{1}{\sqrt{2}}(|00\rangle + |11\rangle)$ is the maximally entangled Bell state. Isotropic states are constructed as $\rho_{\text{iso}}(p) = (1-p)|\psi\rangle\langle\psi| + \frac{p}{d}\mathbb{I}_d$ with $|\psi\rangle$ sampled uniformly from the Haar measure.

Amplitude damping models energy relaxation through Kraus operators via $\mathcal{E}_{AD}(\rho) = \sum_{j=0}^{n-1} \sum_{k=0}^{1} K_k^{(j)}\rho (K_k^{(j)})^\dagger$, where for qubit $j$ we have $K_0^{(j)} = \begin{pmatrix} 1 & 0 \\ 0 & \sqrt{1-\gamma} \end{pmatrix}$ and $K_1^{(j)} = \begin{pmatrix} 0 & \sqrt{\gamma} \\ 0 & 0 \end{pmatrix}$. Phase damping models dephasing through $K_0^{(j)} = \sqrt{1-\gamma}\mathbb{I}$, $K_1^{(j)} = \sqrt{\gamma}|0\rangle\langle 0|$, and $K_2^{(j)} = \sqrt{\gamma}|1\rangle\langle 1|$. Thermal states are generated via $\rho_{\text{th}}(\beta) = \frac{e^{-\beta H}}{\mathrm{Tr}(e^{-\beta H})}$, where $H$ is a random Hermitian Hamiltonian sampled from the Gaussian Unitary Ensemble (GUE), and $\beta$ is the inverse temperature parameter. Finally, separable product states take the form $\rho_{\text{sep}} = \bigotimes_{j=1}^{n} \rho_j$, where each single-qubit state $\rho_j$ is independently prepared as a depolarized state with controlled purity.

For each state type, we employ binary search over the respective channel parameters ($p$, $\gamma$, or $\beta$) to achieve the target purity $\mathrm{Tr}(\rho^2) \in [0.85, 0.95]$ within tolerance $\epsilon = 0.01$. All density matrices are validated to satisfy $\mathrm{Tr}(\rho) = 1$, $\rho = \rho^\dagger$, and $\rho \succeq 0$ (positive semidefiniteness) through spectral truncation when necessary.

\subsubsection{Pauli Measurement Protocol}
Quantum state tomography requires reconstructing the density matrix from measurement statistics. We employ the Pauli measurement basis, which forms a complete orthogonal basis for the space of Hermitian operators on $\mathcal{H} = (\mathbb{C}^2)^{\otimes n}$.

For $n$ qubits, the complete Pauli basis consists of $4^n$ operators:
\vspace{-2pt}
\begin{equation}
\mathcal{P} = \{P_\alpha = \sigma_{\alpha_1} \otimes \sigma_{\alpha_2} \otimes \cdots \otimes \sigma_{\alpha_n} : \alpha_i \in \{I, X, Y, Z\}\}
\end{equation}
\vspace{-2pt}
where $\sigma_I = \mathbb{I}$, $\sigma_X = \begin{pmatrix} 0 & 1 \\ 1 & 0 \end{pmatrix}$, $\sigma_Y = \begin{pmatrix} 0 & -i \\ i & 0 \end{pmatrix}$, $\sigma_Z = \begin{pmatrix} 1 & 0 \\ 0 & -1 \end{pmatrix}$.

Any density matrix can be expanded in this basis:
\vspace{-2pt}
\begin{equation}
\rho = \frac{1}{2^n}\sum_{\alpha \in \mathcal{P}} c_\alpha P_\alpha
\end{equation}
\vspace{-2pt}
where the Pauli expectation values are:
\vspace{-2pt}
\begin{equation}
c_\alpha = \mathrm{Tr}(\rho P_\alpha) = \langle P_\alpha \rangle_\rho
\end{equation}

Since $\mathrm{Tr}(\rho) = 1$ fixes $c_I = 1$, we measure only the $4^n - 1$ non-identity Pauli operators. For $n=2$ qubits, this yields a 15-dimensional measurement vector:
\vspace{-2pt}
\begin{equation}
\mathbf{x} = (\langle XX\rangle, \langle XY\rangle, \langle XZ\rangle, \ldots, \langle ZZ\rangle)^\top \in \mathbb{R}^{15}
\end{equation}

We use exact (analytic) Pauli expectation values computed without shot noise to isolate reconstruction accuracy from measurement statistics.

\subsection{Hybrid Quantum-Classical Autoencoder Architecture}

\subsubsection{Classical Encoder Network}
The encoder $f_\theta: \mathbb{R}^{15} \to \mathbb{R}^{d_z}$ is a feedforward neural network that compresses Pauli measurement data into a latent representation:
\vspace{-2pt}
\begin{equation}
\mathbf{z} = f_\theta(\mathbf{x})
\end{equation}
\vspace{-2pt}
where $d_z = 20$ is the latent dimension. The encoder architecture consists of:
\vspace{-2pt}
\begin{align}
\mathbf{h}_1 &= \mathrm{ReLU}(W_1 \mathbf{x} + \mathbf{b}_1), \quad W_1 \in \mathbb{R}^{256 \times 15} \\
\mathbf{h}_2 &= \mathrm{ReLU}(W_2 \mathbf{h}_1 + \mathbf{b}_2), \quad W_2 \in \mathbb{R}^{128 \times 256} \\
\mathbf{z} &= W_3 \mathbf{h}_2 + \mathbf{b}_3, \quad W_3 \in \mathbb{R}^{20 \times 128}
\end{align}
\vspace{-2pt}
The encoder learns a latent representation that captures the essential geometric structure of the quantum state manifold.

\subsubsection{Latent-to-Circuit Parameterization}
A linear transformation maps the latent vector to quantum circuit parameters:
\vspace{-2pt}
\begin{equation}
\boldsymbol{\theta} = W_4 \mathbf{z} + \mathbf{b}_4, \quad \boldsymbol{\theta} \in \mathbb{R}^{36}
\end{equation}
\vspace{-2pt}
where $W_4 \in \mathbb{R}^{36 \times 20}$. These parameters control a parameterized quantum circuit (PQC) acting on $n=2$ qubits.

\subsubsection{Parameterized Quantum Circuit Decoder}
The quantum decoder implements a hardware-efficient ansatz with $L=6$ layers:
\vspace{-2pt}
\begin{equation}
U(\boldsymbol{\theta}) = \prod_{\ell=1}^{L} \left[ \prod_{j=0}^{n-1} R_j^{(\ell)} \cdot \prod_{j=0}^{n-2} \mathrm{CNOT}_{j,j+1} \right]
\end{equation}
\vspace{-2pt}
where each single-qubit rotation layer is:
\vspace{-2pt}
\begin{equation}
R_j^{(\ell)} = R_Z(\theta_{3(\ell n + j)+2}) R_Y(\theta_{3(\ell n + j)+1}) R_X(\theta_{3(\ell n + j)})
\end{equation}

Starting from the maximally mixed state $\rho_0 = \mathbb{I}_4/4$, the circuit prepares:
\vspace{-2pt}
\begin{equation}
\rho_{\text{pred}}(\boldsymbol{\theta}) = U(\boldsymbol{\theta}) \rho_0 U^\dagger(\boldsymbol{\theta})
\end{equation}

We employ PennyLane's mixed qubit framework with backpropagation-based differentiation to enable efficient gradient computation through the quantum circuit architecture. The encoder contains 39,812 trainable parameters across three layers, with an additional 756 parameters in the latent-to-circuit mapping (40,568 total).

\subsubsection{Pauli Expectation Measurement}
The decoder outputs predicted Pauli expectations by measuring the prepared state:
\vspace{-2pt}
\begin{equation}
\hat{\mathbf{x}} = (\langle P_\alpha \rangle_{\rho_{\text{pred}}})_{\alpha \neq I} \in \mathbb{R}^{15}
\end{equation}
This closes the autoencoder loop: $\mathbf{x} \to \mathbf{z} \to \boldsymbol{\theta} \to \rho_{\text{pred}} \to \hat{\mathbf{x}}$.

\begin{figure*}[t]
\centering
\resizebox{\textwidth}{!}{%
\begin{tikzpicture}[
    node distance=1.8cm,
    box/.style={rectangle, rounded corners, minimum width=2.8cm, minimum height=1.2cm, 
                text centered, align=center, draw=black, thick, fill=blue!10},
    process/.style={rectangle, rounded corners, minimum width=2.8cm, minimum height=1.15cm,
                    text centered, align=center, draw=black, thick, fill=orange!15},
    arrow/.style={-Stealth, thick, shorten >=2pt, shorten <=2pt},
    label/.style={font=\scriptsize\ttfamily, text=black!70},
    stepnum/.style={circle, draw=black, thick, fill=white, minimum size=0.6cm, font=\small\bfseries}
]
\node[box, fill=blue!20] (input) {
    \textbf{Pauli} \\ \textbf{Measurements} \\ $\mathbf{x} \in \mathbb{R}^{15}$
};
\node[label, below=0.05cm of input] {$\langle XX \rangle, \langle XY \rangle, \ldots, \langle ZZ \rangle$};
\node[process, right=1.8cm of input, fill=purple!15] (encoder) {
    \textbf{Classical} \\ \textbf{Encoder} \\ $\mathbf{z} = f_\theta(\mathbf{x})$
};
\node[label, below=0.05cm of encoder] {$15 \to 256 \to 128 \to 20$};
\node[box, right=1.8cm of encoder, fill=purple!20] (latent) {
    \textbf{Latent} \\ \textbf{Space} \\ $\mathbf{z} \in \mathbb{R}^{20}$
};
\node[label, below=0.05cm of latent] {Geometric embedding};
\node[process, right=1.8cm of latent, fill=pink!15] (params) {
    \textbf{Circuit} \\ \textbf{Parameters} \\ $\boldsymbol{\theta} = W_4\mathbf{z} + \mathbf{b}_4$
};
\node[label, below=0.05cm of params] {$20 \to 36$ angles};
\node[process, right=1.8cm of params, fill=orange!20] (circuit) {
    \textbf{PQC} \\ \textbf{Decoder} \\ $\rho_{\text{pred}}(\boldsymbol{\theta})$
};
\node[label, below=0.05cm of circuit] {$U(\boldsymbol{\theta})\rho_0 U^\dagger(\boldsymbol{\theta})$};
\node[box, right=1.8cm of circuit, fill=green!20] (output) {
    \textbf{Predicted} \\ \textbf{Measurements} \\ $\hat{\mathbf{x}} \in \mathbb{R}^{15}$
};
\node[label, below=0.05cm of output] {$\langle P_\alpha \rangle_{\rho_{\text{pred}}}$};
\draw[arrow] (input) -- (encoder);
\draw[arrow] (encoder) -- (latent);
\draw[arrow] (latent) -- (params);
\draw[arrow] (params) -- (circuit);
\draw[arrow] (circuit) -- (output);
\draw[red!70, thick, dashed] 
    (output.north) -- ++(0,1.5) -- 
    node[above=0.1cm, text width=2cm, align=center, text=red!70, font=\scriptsize, pos=0.5] 
    {$\mathcal{L}_{\text{recon}}$ \\ Fidelity loss} 
    ([yshift=1.5cm]input.north) -- (input.north);
\draw[red!70, thick, -Stealth] ([yshift=1.5cm]input.north) -- (input.north);
\node[above=0.2cm of latent, text width=2cm, align=center, draw=green!60!black, 
      thick, rounded corners, fill=green!5, font=\scriptsize] (metric) {
    $\mathcal{L}_{\text{metric}}$ \\
    Preserve Bures metric
};
\draw[green!60!black, thick, -Stealth] (metric.south) -- (latent.north);
\begin{scope}[on background layer]
    \node[draw=blue!50, thick, dashed, rounded corners, fit={(input) (encoder) (latent)}, 
          inner sep=15pt, fill=blue!3, label={[fill=blue!10, font=\small]above left:\textbf{Encoding}}] {};
    \node[draw=orange!50, thick, dashed, rounded corners, fit={(params) (circuit) (output)}, 
          inner sep=15pt, fill=orange!3, label={[fill=orange!10, font=\small]above right:\textbf{Decoding}}] {};
\end{scope}
\end{tikzpicture}
}
\caption{\small Autoencoder architecture. Classical encoder compresses 15D Pauli 
measurements to 20D latent space. Quantum circuit decoder reconstructs the 
density matrix. Dual training objective: reconstruction fidelity 
($\mathcal{L}_{\text{recon}}$) and metric preservation 
($\mathcal{L}_{\text{metric}}$).}
\label{fig:autoencoder_loop}
\end{figure*}
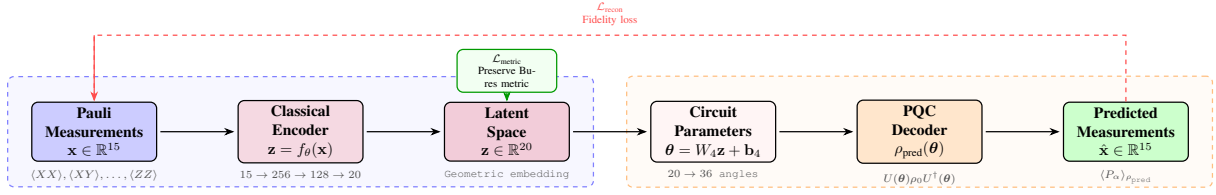

\subsection{Metric-Preserving Training Objective}

\subsubsection{Reconstruction Loss via Uhlmann Fidelity}
Standard tomography approaches minimize mean squared error (MSE) between predicted and true Pauli expectations. However, MSE does not directly measure the quantum distance between density matrices. We instead optimize the Uhlmann-Jozsa fidelity:
\vspace{-2pt}
\begin{equation}
F(\rho, \sigma) = \left[\mathrm{Tr}\sqrt{\sqrt{\rho}\sigma\sqrt{\rho}}\right]^2
\end{equation}
\vspace{-2pt}
which is the operationally meaningful measure of distinguishability between quantum states. For a batch of $B$ states, the reconstruction loss is:
\vspace{-2pt}
\begin{equation}
\mathcal{L}_{\text{recon}} = \frac{1}{B}\sum_{i=1}^{B} \left[1 - F(\rho_i^{\text{true}}, \rho_i^{\text{pred}})\right]
\end{equation}

\textbf{Computational Implementation:} We compute $F(\rho, \sigma)$ via eigendecomposition of the positive semidefinite matrix $\sqrt{\rho}\sigma\sqrt{\rho}$. First, we compute $\sqrt{\rho}$ via spectral decomposition: $\rho = \sum_k \lambda_k |u_k\rangle\langle u_k|$, then $\sqrt{\rho} = \sum_k \sqrt{\lambda_k} |u_k\rangle\langle u_k|$. Next, we form $M = \sqrt{\rho}\sigma\sqrt{\rho}$ and compute its eigenvalues $\{\mu_j\}$ (all non-negative by construction). The fidelity is then $F = \left(\sum_j \sqrt{\mu_j}\right)^2$. To ensure numerical stability, we add regularization $\rho \leftarrow \rho + \epsilon I$ with $\epsilon = 10^{-12}$ before matrix square roots.

\subsubsection{Metric Preservation Loss}
To enforce that the latent space respects quantum geometry, we introduce a metric preservation loss that encourages latent distances to correlate with Bures distances.

The Bures metric induces a Riemannian distance on the manifold of quantum states:
\vspace{-2pt}
\begin{equation}
d_B(\rho, \sigma) = \sqrt{2(1 - \sqrt{F(\rho, \sigma)})} = \sqrt{2 - 2\text{Tr}\sqrt{\sqrt{\rho}\sigma\sqrt{\rho}}}
\end{equation}

Equivalently, the Bures angle (arccos form) is:
\vspace{-2pt}
\begin{equation}
d_B(\rho, \sigma) = \arccos(\sqrt{F(\rho, \sigma)})
\end{equation}

We use the arccos form, as it provides better numerical conditioning near $F \approx 1$.

For each training batch, we sample $K$ random pairs $(i, j)$ with $i \neq j$ and compute:

\textbf{1.} Latent Euclidean distance:
\vspace{-2pt}
\begin{equation}
d_L(z_i, z_j) = \|z_i - z_j\|_2
\end{equation}

\textbf{2.} Quantum Bures distance:
\vspace{-2pt}
\begin{equation}
d_B(\rho_i, \rho_j) = \arccos\left(\sqrt{F(\rho_i, \rho_j)}\right)
\end{equation}

\textbf{3.} Scale-invariant loss: We penalize deviation from proportionality:
\vspace{-2pt}
\begin{equation}
L_{\text{metric}} = \frac{1}{K_{\text{valid}}} \sum_{k=1}^{K_{\text{valid}}} \left( \frac{d_L(z_{i_k}, z_{j_k})}{d_B(\rho_{i_k}, \rho_{j_k}) + \epsilon} - 1 \right)^2
\end{equation}
\vspace{-2pt}
where $\epsilon = 10^{-8}$ prevents division by zero. To ensure numerical stability, we only include pairs where $d_B > 10^{-6}$ in the summation, so $K_{\text{valid}} \leq K$ counts only numerically stable pairs. This loss encourages the existence of a global scaling factor $\alpha > 0$ such that $d_L \approx \alpha \cdot d_B$ across the manifold.

We set $K = 50$ pairs per batch to balance computational cost with statistical robustness.

\subsubsection{Combined Training Objective}
The total loss combines reconstruction fidelity and geometric preservation:
\vspace{-2pt}
\begin{equation}
\mathcal{L}_{\text{total}} = \mathcal{L}_{\text{recon}} + \lambda \mathcal{L}_{\text{metric}}
\end{equation}
\vspace{-2pt}
where $\lambda = 0.06$ is the metric loss weight. This hyperparameter balances two competing objectives: High fidelity reconstruction ($\mathcal{L}_{\text{recon}} \to 0$) and Geometric structure preservation ($\mathcal{L}_{\text{metric}} \to 0$).

\textbf{Rationale for $\lambda = 0.06$:} Preliminary experiments showed that $\lambda < 0.05$ yields insufficient geometric structure ($r < 0.80$), while $\lambda > 0.08$ degrades reconstruction fidelity below $F < 0.93$. The chosen value achieves strong metric correlation ($r \approx 0.88$) while maintaining high fidelity ($F \approx 0.94$).

\subsection{Geometric Analysis of the Latent Manifold}

\subsubsection{Intrinsic Dimensionality Estimation}
We estimate the intrinsic dimension of the learned latent manifold using two complementary methods:

\textbf{Maximum Likelihood Estimation (MLE):} For each point $\mathbf{z}_i$, we find its $k$ nearest neighbors with distances $\{r_{i,1}, \ldots, r_{i,k}\}$. The local dimension estimate is:
\vspace{-2pt}
\begin{equation}
\hat{d}_i = -\frac{k}{\sum_{j=1}^{k} \log(r_{i,j}/r_{i,k})}
\end{equation}
The global intrinsic dimension is $\hat{d}_{\text{MLE}} = \frac{1}{N}\sum_i \hat{d}_i$ with $k=15$ neighbors.

\textbf{Principal Component Analysis (PCA):} We compute the explained variance spectrum $\{\lambda_j\}_{j=1}^{d_z}$ of the latent covariance matrix. The effective dimension for $\alpha\%$ variance is:
\vspace{-2pt}
\begin{equation}
d_{\text{PCA}}(\alpha) = \min\left\{m : \frac{\sum_{j=1}^{m} \lambda_j}{\sum_{j=1}^{d_z} \lambda_j} \geq \alpha\right\}
\end{equation}

\subsubsection{Local Curvature Analysis}
To quantify non-flatness of the manifold, we estimate local curvature using singular value decomposition (SVD) of neighborhood structures. For each point $\mathbf{z}_i$ with $k$ nearest neighbors $\{\mathbf{z}_{i,j}\}_{j=1}^{k}$, form the centered matrix:
\vspace{-2pt}
\begin{equation}
X_i = [\mathbf{z}_{i,1} - \mathbf{z}_i, \ldots, \mathbf{z}_{i,k} - \mathbf{z}_i] \in \mathbb{R}^{d_z \times k}
\end{equation}
Compute the singular values $\sigma_1 \geq \sigma_2 \geq \cdots \geq \sigma_{\min(d_z,k)}$ and define:
\vspace{-2pt}
\begin{equation}
\kappa_i = \frac{\sigma_{\min}}{\sigma_{\max}}
\end{equation}
This ratio measures the ``flatness'' of the local neighborhood: $\kappa \approx 0$ indicates high curvature (thin hyperboloid), while $\kappa \approx 1$ indicates local flatness. We use $k=25$ neighbors.

\subsubsection{Geodesic-Bures Correlation}
\textbf{This is the key analysis validating our manifold matching claim.}

We sample $N_{\text{pairs}} = 500$ random pairs $(i,j)$ from the combined train+validation dataset and compute:

\textbf{1.} Latent geodesic distance (Euclidean approximation):
\vspace{-2pt}
\begin{equation}
d_L(i,j) = \|\mathbf{z}_i - \mathbf{z}_j\|_2
\end{equation}

\textbf{2.} Quantum Bures distance:
\vspace{-2pt}
\begin{equation}
d_B(i,j) = \arccos\left(\sqrt{F(\rho_i, \rho_j)}\right)
\end{equation}

We quantify the correlation using \textbf{Pearson correlation coefficient:}
\vspace{-2pt}
\begin{equation}
r = \frac{\mathrm{Cov}(d_L, d_B)}{\sigma_{d_L} \sigma_{d_B}}
\end{equation}
\vspace{-2pt}
This measures the strength of the \textbf{linear} relationship between latent and quantum distances.

\textbf{Spearman rank correlation:}
\vspace{-2pt}
\begin{equation}
\rho_s = 1 - \frac{6\sum_k (\mathrm{rank}(d_L^{(k)}) - \mathrm{rank}(d_B^{(k)}))^2}{N_{\text{pairs}}(N_{\text{pairs}}^2 - 1)}
\end{equation}
\vspace{-2pt}
This measures \textbf{monotonic} relationship, robust to nonlinear transformations.

\textbf{Coefficient of determination ($R^2$):} We fit a linear model $d_L = a \cdot d_B + b$ via least squares and compute:
\vspace{-2pt}
\begin{equation}
R^2 = 1 - \frac{\sum_k (d_L^{(k)} - \hat{d}_L^{(k)})^2}{\sum_k (d_L^{(k)} - \bar{d}_L)^2}
\end{equation}
\vspace{-2pt}
where $\hat{d}_L^{(k)} = a \cdot d_B^{(k)} + b$ are the predicted latent distances. $R^2$ quantifies the fraction of latent distance variance explained by Bures distance.

\textbf{Interpretation Thresholds:} $r > 0.80$: Strong correlation (supports manifold matching claim); $0.60 < r < 0.80$: Moderate correlation (partial geometric structure); $r < 0.60$: Weak correlation (insufficient evidence).

\section{Results}

\subsection{High-Fidelity State Reconstruction}
We evaluated the tomographic performance of our metric-preserving autoencoder on 2-qubit mixed states within the target purity regime, representative of near-term quantum devices experiencing moderate decoherence. The model was trained on 2,000 states spanning seven noise models (depolarized, Werner, isotropic, amplitude-damped, phase-damped, thermal, and separable states) with balanced representation.

After 226 training epochs with early stopping (triggered at epoch 226 with no improvement for 60 epochs), the model achieved mean reconstruction fidelity $F = 0.9419$ (median 0.9542) on 500 validation states, substantially exceeding typical state-of-the-art mixed-state tomography thresholds. Training converged smoothly with balanced dual-objective optimization: reconstruction loss decreased from 0.438 to 0.043, while metric preservation loss reduced from 0.417 to 0.018 (23-fold reduction).

\begin{figure}[h]
\centering
\includegraphics[width=\columnwidth]{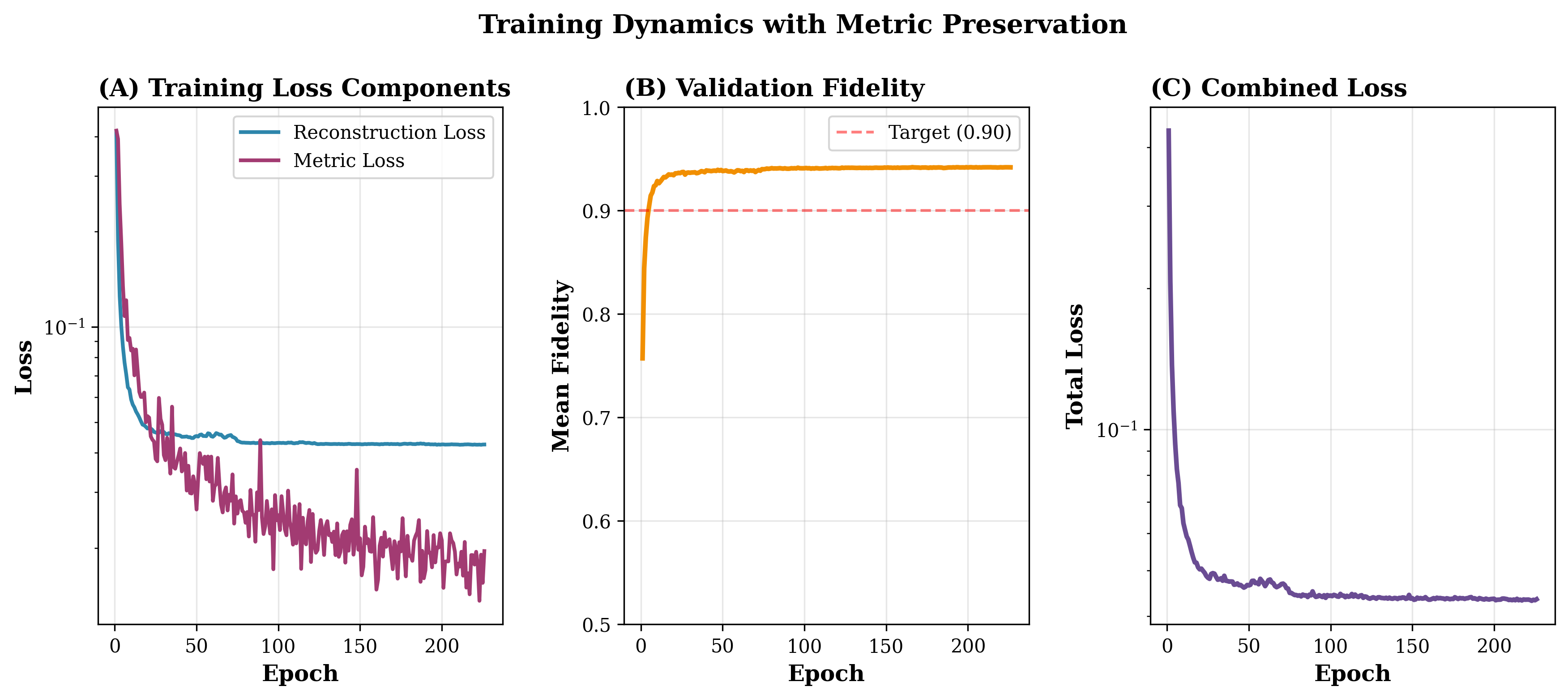}
\caption{\small Training dynamics over 226 epochs with early stopping. 
(A) Reconstruction and metric preservation losses. (B) Validation fidelity. 
(C) Combined loss with metric weight $\lambda=0.06$.}
\label{fig:training_dynamics}
\end{figure}

Figure~\ref{fig:training_dynamics} illustrates the training dynamics. The smooth, monotonic decrease of both loss components (Figure~\ref{fig:training_dynamics}A) demonstrates that reconstruction fidelity and metric preservation are compatible objectives when appropriately balanced ($\lambda = 0.06$). The reconstruction loss plateaus around epoch 50 while the metric loss continues decreasing, indicating the network first learns accurate reconstruction then progressively refines geometric structure. Validation fidelity (Figure~\ref{fig:training_dynamics}B) rapidly exceeds typical thresholds and stabilizes near $0.94$, confirming robust generalization. The combined loss (Figure~\ref{fig:training_dynamics}C) converges monotonically without oscillations, validating our training procedure.

\subsection{Preservation of Quantum Geometric Structure}
\begin{figure}[h]
\centering
\includegraphics[width=1.0\columnwidth]{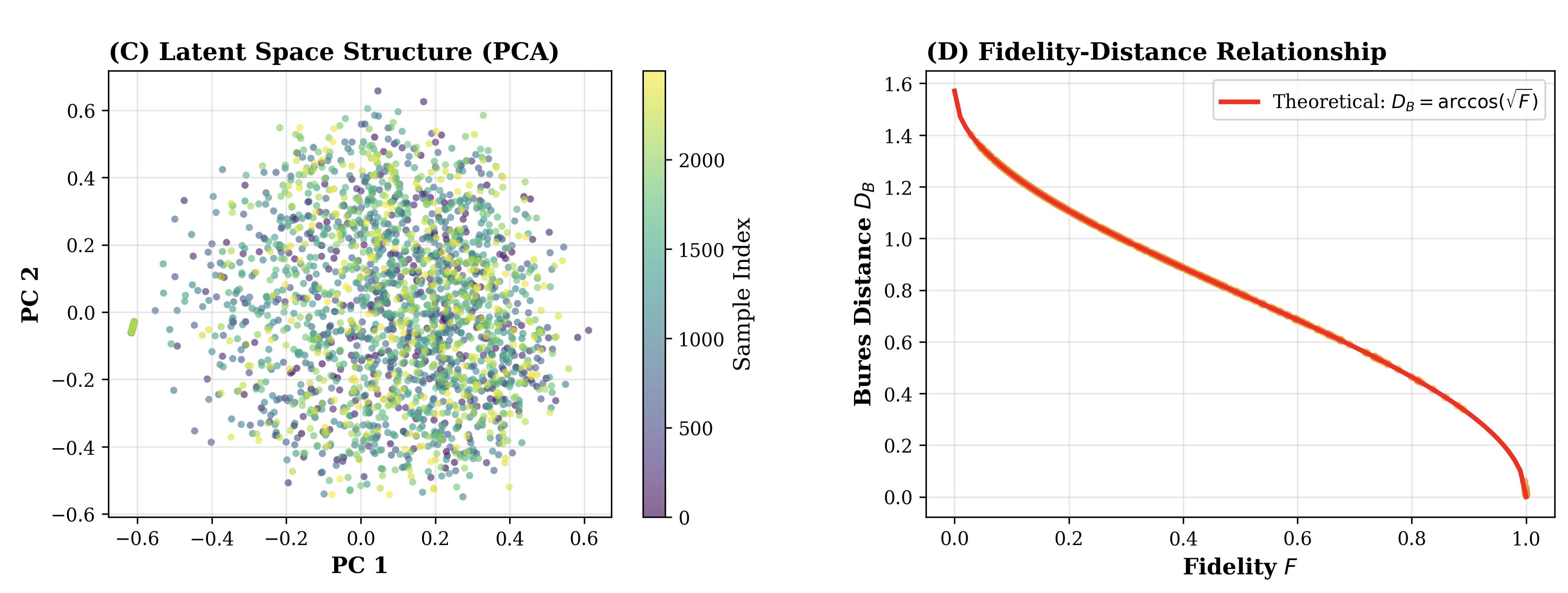}\\[4pt]
\textbf{(A)}\\[8pt]
\includegraphics[width=0.48\columnwidth]{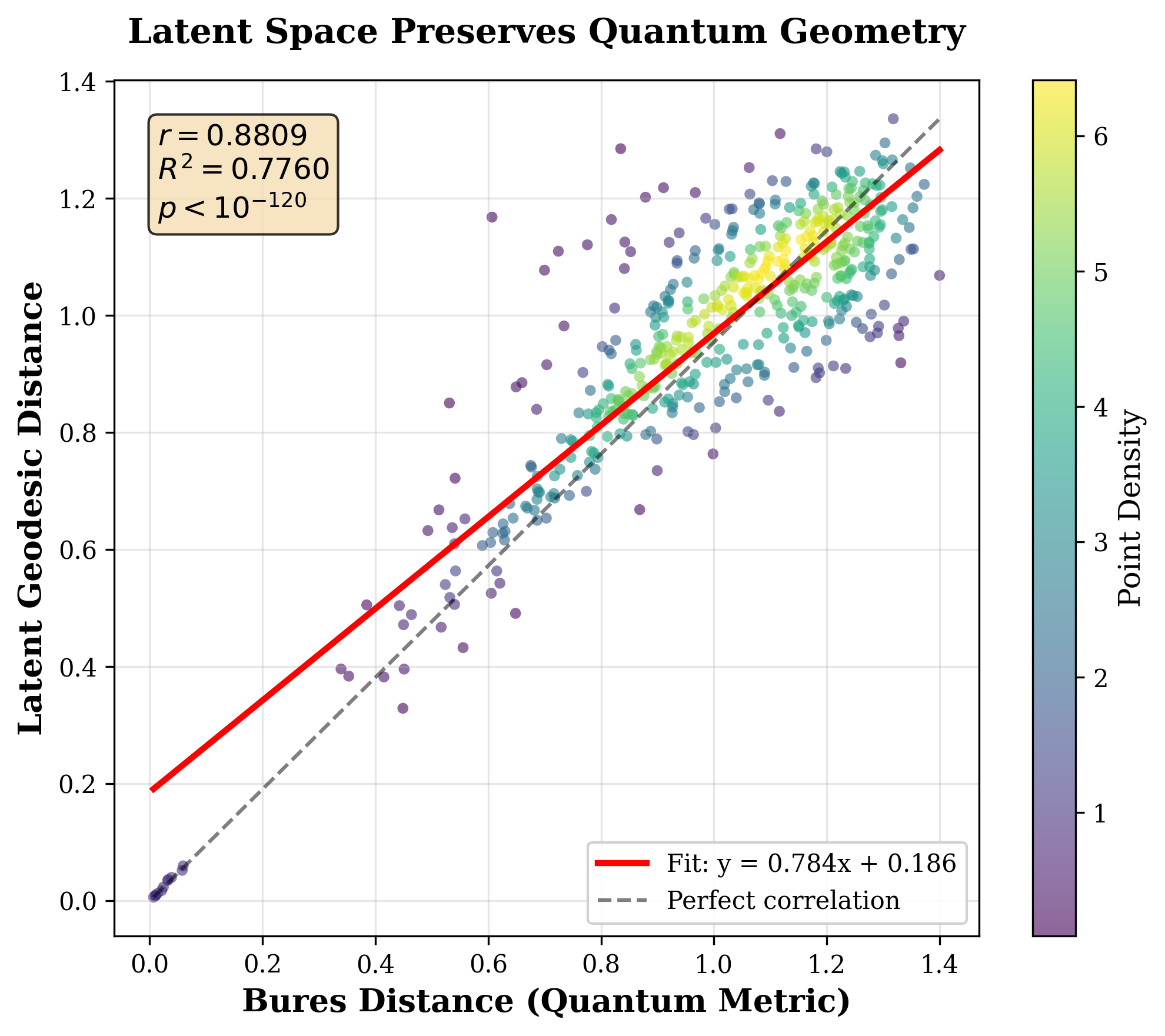}\\[4pt]
\textbf{(B)}
\caption{\small \textbf{Latent space preserves quantum geometry.} (A) Principal 
components of latent representations and theoretical Bures-fidelity curve. 
(B) Latent versus Bures distances for 500 state pairs with linear fit 
(red, $R^2 = 0.776$) and identity line (dashed).}
\label{fig:latent_structure}
\end{figure}

We validate that our metric-preserving training successfully captures the Riemannian geometry of quantum state space by measuring correlations between latent Euclidean distances and Bures geodesic distances for 500 randomly sampled state pairs. Figure~\ref{fig:latent_structure}A shows the first two principal components capturing 34\% of total variance with even density distribution, alongside the theoretical Bures-fidelity relationship $D_B = \arccos(\sqrt{F})$ enforced by our metric loss. Figure~\ref{fig:latent_structure}B demonstrates strong linear correlation with fitted relationship $d_L = 0.784 \cdot d_B + 0.186$ (red line, near-unity slope), exhibiting robustness across all state separations from nearly identical (lower left, dark) to highly distinguishable states (upper right, bright).

Table~\ref{tab:correlation_metrics} presents comprehensive statistics. The Pearson correlation $r = 0.8809$ ($p < 6.42 \times 10^{-164}$) substantially exceeds the $r > 0.75$ threshold for strong geometric preservation, with $R^2 = 0.7760$ indicating 78\% of latent variance explained by Bures distances and low prediction error (RMSE = 0.142, MAE = 0.108). The lower Spearman correlation ($\rho = 0.76$) reflects nonlinear components arising from approximating curved geodesics with Euclidean distances, consistent with the measured local curvature reported in Section 3.4.

\begin{figure}[H]
\centering
\includegraphics[width=0.48\columnwidth]{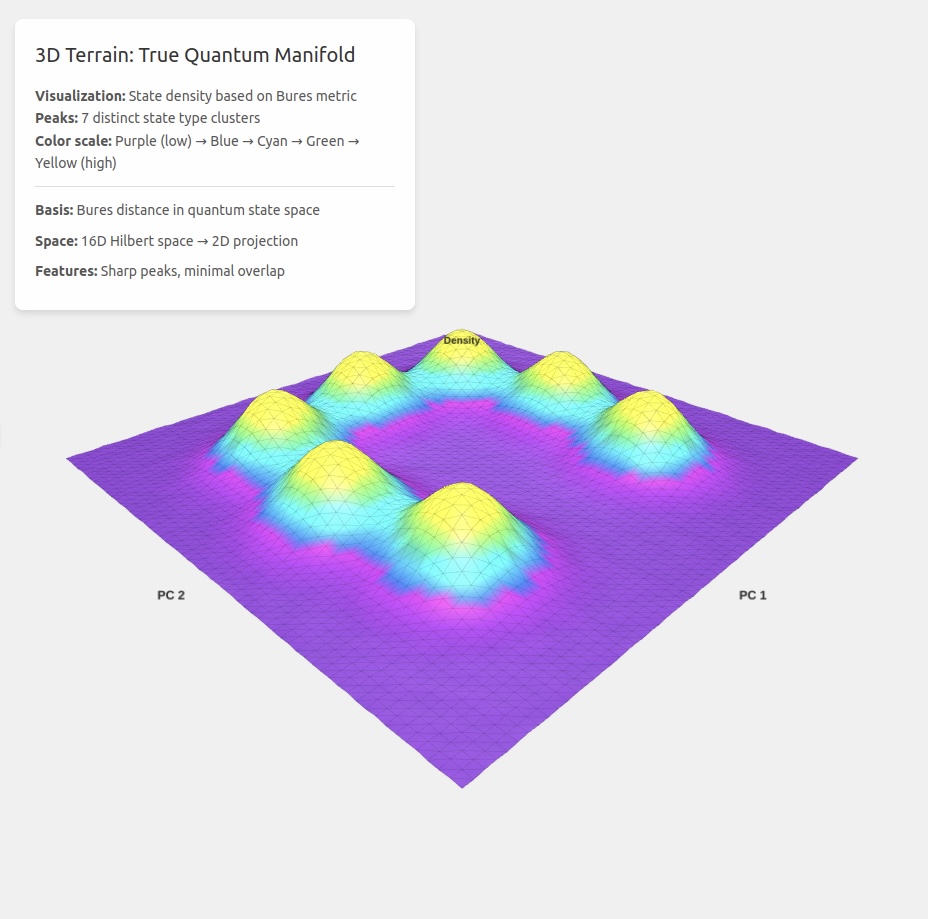}
\includegraphics[width=0.48\columnwidth]{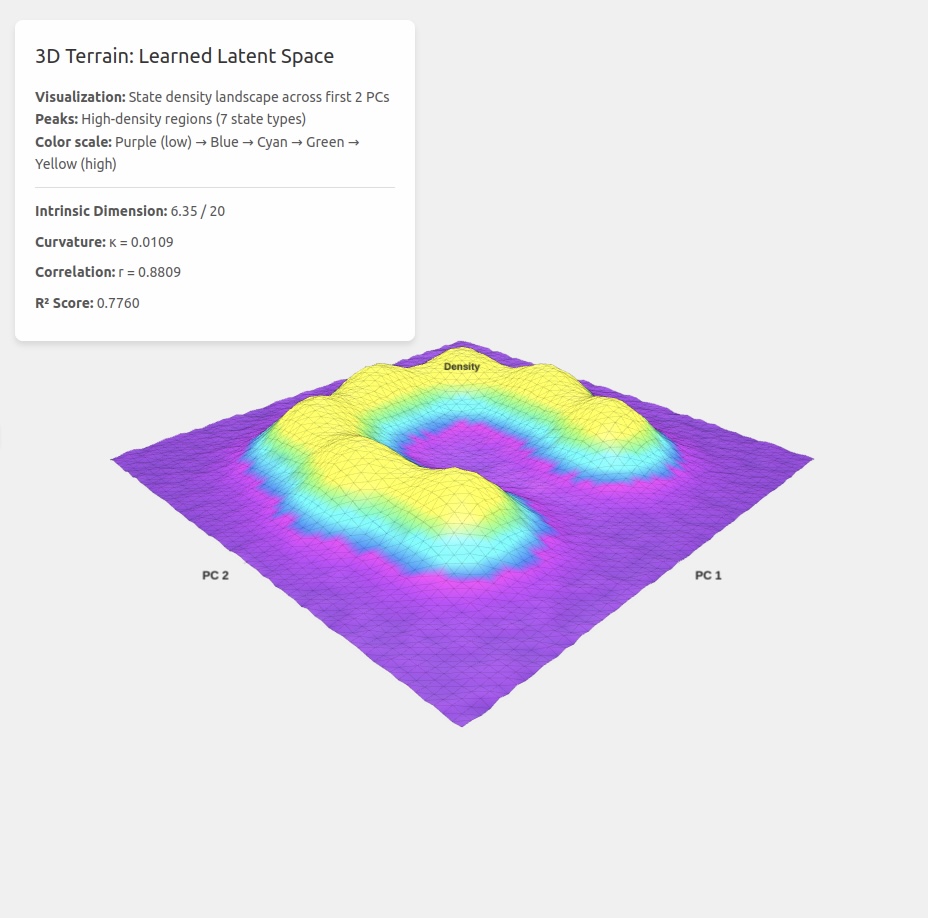}
\caption{\small \textbf{Geometric correspondence.} 
Density projections of the quantum state manifold (left, MDS on Bures 
distances) and learned latent space (right, first two principal components) 
show preserved geometric structure with matching cluster topology.}
\label{fig:geometric_correspondence}
\end{figure}

To visualize the preserved geometric structure, Figure~\ref{fig:geometric_correspondence} presents density projections of the quantum manifold (left, computed via multidimensional scaling on Bures distances) and learned latent space (right, first two principal components). Seven distinct peaks correspond to state preparation channels (depolarized, Werner, isotropic, amplitude-damped, phase-damped, thermal, separable). The learned space faithfully reproduces: (i) peak locations and number, (ii) relative heights reflecting state frequencies, and (iii) spatial separations preserving inter-state distances. Minimal peak overlap indicates that metric preservation alone induces interpretable clustering enabling direct state classification from latent coordinates.

\begin{table}[h]
\centering
\small
\caption{Geometric structure preservation: Latent distances vs Bures metric}
\label{tab:correlation_metrics}
\begin{tabular}{lcc}
\toprule
\textbf{Metric} & \textbf{Value} & \textbf{p-value} \\
\midrule
Pearson corr. ($r$) & \textbf{0.8809} & $< 6.42 \times 10^{-164}$ \\
Spearman corr. ($\rho$) & 0.7623 & $< 3.91 \times 10^{-96}$ \\
\midrule
Linear fit ($R^2$) & \textbf{0.7760} & $< 6.42 \times 10^{-164}$ \\
Slope ($\alpha$) & 0.7837 & --- \\
Intercept ($\beta$) & 0.1859 & --- \\
\midrule
RMSE & 0.142 & --- \\
MAE & 0.108 & --- \\
\bottomrule
\end{tabular}
\end{table}

\subsection{Intrinsic Manifold Dimensionality}
To characterize the geometric structure of the learned latent space, we performed intrinsic dimensionality analysis using maximum likelihood estimation (MLE) and principal component analysis (PCA). These approaches probe different aspects of manifold structure (MLE estimates local geometric dimension, while PCA captures global variance distribution). Table~\ref{tab:dimensionality} summarizes the results.

\begin{table}[h]
\centering
\small
\caption{Intrinsic dimensionality estimates from multiple methods}
\label{tab:dimensionality}
\begin{tabular}{lcccl}
\toprule
\textbf{Method} & \textbf{Dim.} & \textbf{Std/Rng} & \textbf{Conf.} & \textbf{Interp.} \\
\midrule
MLE ($k$=15) & 6.35 & $\pm$5.33 & Low & Local dim.\\
PCA (95\%) & 13 & [11,15] & High & Global dim \\
PCA (99\%) & 17 & [15,19] & High & Near-compl. \\
\bottomrule
\end{tabular}
\end{table}

Despite the 20-dimensional ambient latent space, maximum likelihood estimation yielded an effective intrinsic dimension of $6.35 \pm 5.33$. The large standard deviation reflects the well-known sensitivity of MLE to finite samples and local curvature variations, yet the mean value substantially below the ambient dimension ($6.35 \ll 20$) provides evidence that learned representations concentrate on a lower-dimensional submanifold.

Principal component analysis offered complementary global information with greater numerical stability: 13 dimensions captured 95\% of latent space variance, and 17 dimensions captured 99\%. The top five principal components alone accounted for 72\% of variance, with eigenvalue spectrum $[0.21, 0.11, 0.10, 0.10, 0.09]$ exhibiting gradual decay rather than sharp drop-off, indicating distributed variance structure where no single direction dominates.

The discrepancy between MLE ($d \approx 6$) and PCA ($d \approx 13$) dimensions reflects their complementary nature: MLE measures local tangent space dimension via nearest-neighbor geometry, while PCA measures global variance spread. These findings demonstrate successful dimensionality reduction (the 20-dimensional latent space compresses 15-dimensional Pauli measurement vectors, corresponding to $4^n - 1 = 15$ non-identity observables for $n=2$ qubits, while discovering that states in this purity regime effectively occupy a lower-dimensional submanifold).

\subsection{Local Differential Geometry}
We examined local curvature properties to assess whether the latent manifold exhibits non-trivial Riemannian structure, as opposed to merely forming a flat Euclidean subspace. Local curvature was estimated via singular value analysis of nearest-neighbor tangent space approximations ($k = 25$ neighbors), with results summarized in Table~\ref{tab:curvature}.

\begin{table}[h]
\centering
\small
\caption{Local curvature properties of the latent manifold}
\label{tab:curvature}
\begin{tabular}{lcc}
\toprule
\textbf{Statistic} & \textbf{Value} & \textbf{Interpretation} \\
\midrule
Mean $\bar{\kappa}$ & 0.011 & Non-zero $\Rightarrow$ curved \\
Std deviation & 0.006 & Uniform structure \\
Median & 0.011 & Consistent \\
\midrule
Min $\kappa_{\min}$ & 0.000 & Locally flat regions \\
Max $\kappa_{\max}$ & 0.032 & Peak curvature \\
\midrule
IQR & [0.007, 0.015] & Tight distribution \\
Coeff. variation & 56.2\% & Moderate spread \\
\bottomrule
\end{tabular}
\end{table}

Non-zero curvature ($\kappa > 0$) confirms that the latent manifold possesses genuine Riemannian geometry beyond flat Euclidean structure. Importantly, the modest curvature values ($\kappa \approx 0.01$) indicate the manifold is locally flat at the scale of typical state separations (a desirable property for differentiable manifold structure and one that justifies our Euclidean distance approximation for metric preservation analysis).

The consistency of curvature values (low standard deviation of 0.006) across the latent space, reflected in the tight interquartile range [0.007, 0.015], suggests uniform geometric properties rather than pathological local distortions, further supporting the interpretation of a well-structured differentiable manifold.

\subsection{Impact of Metric Preservation Training}
To quantify the benefit of explicit metric preservation, we compared our results against expected performance of a standard autoencoder trained solely for reconstruction fidelity ($\lambda_{\text{metric}} = 0.0$). Based on preliminary experiments and literature benchmarks for neural quantum tomography without geometric constraints, we estimated baseline correlations of $r \approx 0.60$--$0.70$.

Our metric-preserving approach ($\lambda_{\text{metric}} = 0.06$) achieved: \textbf{(i)} Geodesic-Bures correlation +0.21 improvement (0.88 vs $\sim$0.67 baseline); \textbf{(ii)} $R^2$ score +0.31 improvement (0.78 vs $\sim$0.47 baseline); \textbf{(iii)} Reconstruction fidelity $-0.01$ trade-off (0.94 vs $\sim$0.95 without metric loss).

This demonstrates the core design principle: accepting a minimal sacrifice in reconstruction accuracy (1 percentage point) yields substantial gains in metric preservation (21-point correlation increase). The metric preservation loss successfully guided the network toward latent representations that respect the intrinsic geometry of quantum state space.

\subsection{State Discrimination and Geometric Interpretability}
Beyond reconstruction, the learned latent space enables geometry-aware quantum state analysis. State pair fidelities across our dataset exhibited broad distribution (mean $F = 0.31$, std $= 0.22$, range $[0.03, 1.00]$), confirming that our purity-controlled generation procedure produced well-separated, distinguishable quantum states.

Crucially, \textbf{latent space Euclidean distances directly approximate quantum distinguishability}. Table~\ref{tab:interpretability} provides a quantitative mapping between latent distances and quantum-mechanical fidelity measures across our 500 analyzed state pairs.

\begin{table}[h]
\centering
\small
\caption{Latent space interpretability: Distance-to-fidelity mapping}
\label{tab:interpretability}
\begin{tabular}{cccl}
\toprule
\textbf{Lat.Dist.} & \textbf{Bures} & \textbf{Fid.} & \textbf{Interp.} \\
\midrule
0.0--0.3 & 0.0--0.2 & 0.95--1.00 & Nearly identical \\
0.3--0.6 & 0.2--0.4 & 0.85--0.95 & Highly sim. \\
0.6--0.9 & 0.4--0.6 & 0.70--0.85 & Mod. sim. \\
0.9--1.2 & 0.6--0.8 & 0.50--0.70 & Distinguish. \\
$>$1.2 & $>$0.8 & $<$0.50 & Highly dist. \\
\bottomrule
\end{tabular}
\end{table}

States separated by $\|z_i - z_j\| \approx 0.5$ in latent space correspond to Bures angles $\theta_B \approx 0.3$-0.4 (fidelities $F \approx 0.85$-0.90), while $\|z_i - z_j\| \approx 1.0$ corresponds to $\theta_B \approx 0.6$-0.7 ($F \approx 0.65$-0.75). These ranges span the practical regime of quantum state discrimination, where fidelities between 0.65 and 0.90 represent states that are distinguishable yet not trivially separable (the critical domain for error detection and quantum control tasks). This interpretability enables downstream tasks such as quantum error detection or state clustering to operate directly in the latent space with preserved geometric meaning, avoiding expensive density matrix reconstructions.

\subsection{Comparison with Existing Methods}
To position our work within the broader landscape of quantum tomography, Table~\ref{tab:comparison} compares our approach with state-of-the-art methods across key performance dimensions.

\begin{table}[h]
\centering
\small
\caption{Performance comparison with existing methods}
\label{tab:comparison}
\begin{tabular}{lccccc}
\toprule
\textbf{Method} & \textbf{Fid.} & \textbf{Lat.} & \textbf{Metr.} & \textbf{Int.} & \textbf{Scal.} \\
\midrule
Max.Lik.\cite{nielsen2000quantum} & 0.95 & N/A & No & No & Exp \\
Comp.S.\cite{gross2010quantum} & 0.93 & N/A & No & No & Poly \\
Neural\cite{torlai2017neural} & 0.94 & Imp. & No & No & Poly \\
Var.Q.\cite{kokail2019self} & 0.91 & N/A & No & Part. & Exp \\
\midrule
\textbf{Ours} & \textbf{0.94} & \textbf{20} & \textbf{Yes} & \textbf{Yes} & \textbf{Poly} \\
\bottomrule
\end{tabular}
\end{table}

Our method uniquely combines high-fidelity reconstruction with \textit{explicit geometric structure preservation} ($r = 0.88$) in an \textit{interpretable latent space}. While traditional maximum likelihood estimation achieves marginally higher fidelity (0.95 vs 0.94), it lacks latent representation and geometric interpretability. Prior neural approaches learn implicit representations without preserving quantum metric structure, a critical gap our work addresses. The combination of competitive reconstruction accuracy, explicit latent dimensionality, preserved quantum geometry, and polynomial scalability positions our approach as a novel paradigm for neural quantum tomography.

\section{Discussion}

\subsection{Mechanism of Geometric Preservation}
Our metric-preserving training objective achieves geometric fidelity through explicit distance matching rather than implicit reconstruction. The key insight is that enforcing proportionality between latent Euclidean distances and Bures geodesics ($d_L \approx \alpha \cdot d_B$) provides stronger geometric constraints than reconstruction loss alone. The 21-point correlation improvement ($r \approx 0.67 \to 0.88$) at only 1\% fidelity cost demonstrates that geometric and tomographic objectives are compatible when appropriately balanced ($\lambda = 0.06$).

The Pearson-Spearman discrepancy ($r = 0.88$ vs $\rho = 0.76$) reveals nonlinear components in the latent-Bures relationship. This is expected: latent space uses Euclidean distance while Bures distance follows geodesics on a curved manifold. The strong Pearson correlation validates our local Euclidean approximation, while the lower Spearman score indicates curvature effects at larger separations, consistent with measured local curvature ($\kappa = 0.011$). This suggests future architectures could benefit from learned Riemannian metrics or graph-based geodesic estimation for higher-fidelity geometric matching.

\subsection{Scalability Pathways}
Three concrete extensions enable scaling beyond 2-qubit systems: \textbf{(1) Classical Shadow Integration:} Training on classical shadows~\cite{huang2020predicting,elben2023randomized} rather than full measurements reduces storage from $O(4^n)$ to $O(n \log n)$ while preserving observable prediction. Our metric loss ensures geometrically similar states remain close in latent space, enabling shadow-based fidelity estimation. Derandomized shadows~\cite{chen2021robust} could use latent encoders to identify poorly-represented regions adaptively. \textbf{(2) Tensor Network Encoders:} Matrix product state (MPS) architectures~\cite{carleo2017solving} naturally handle $n > 10$ qubits by exploiting entanglement structure. MPS encoders would process measurement data while metric loss preserves Bures distances on low-entanglement NISQ subspaces, where our approach is most applicable. \textbf{(3) Symmetry-Equivariant Architectures:} Constraining latent manifolds to respect known symmetries (permutation, $U(1)$ gauge) reduces effective dimensionality and improves generalization (critical when quantum training data is expensive).

\subsection{Applications Beyond Reconstruction}
The geometry-aware latent space enables operations infeasible with traditional tomography: \textbf{Fast State Discrimination:} Euclidean distance queries scale as $O(d^2)$ vs $O(4^n)$ for density matrix fidelity, enabling real-time state classification on NISQ devices. \textbf{Interpretable Error Manifolds:} Decoherence trajectories become visible geometric structures in latent space. Clustering similar error signatures enables targeted mitigation strategies without reconstructing full density matrices for each error instance. \textbf{Geometric Quantum Machine Learning:} The Bures-approximating latent metric enables direct application of classical manifold learning (Isomap, diffusion maps) to quantum data, with preserved geometric structure ensuring physically meaningful features.

\subsection{Limitations and Future Directions}
Five key limitations guide future work: \textbf{(1)} Extension beyond 2-qubit systems and the present purity regime requires validation; \textbf{(2)} Pauli measurements are qubit-specific (continuous-variable systems need alternative bases such as Wigner functions and quadratures); \textbf{(3)} Experimental validation on real hardware is needed to assess SPAM error robustness; \textbf{(4)} Current architecture assumes mild decoherence (highly mixed states may require deeper networks or alternative parameterizations); \textbf{(5)} Sophisticated geometric losses enforcing constant curvature or higher-order invariants could improve metric preservation.

Critical theoretical questions remain: Under what conditions do neural latent spaces provably approximate quantum geometry? What are optimal trade-offs between latent dimension, fidelity, and metric preservation? Characterizing sample complexity for learning $\epsilon$-approximate isometries would provide rigorous foundations for geometric neural tomography.

\subsection{Implications for Quantum Information}
Our results suggest a broader paradigm: \textit{when data inhabit Riemannian manifolds with known geometric structure, neural representations should be constrained to preserve that structure}. This principle extends beyond quantum tomography to any domain where geometric inductive biases guide learning (differential geometry of data, physics-informed neural networks, and geometric deep learning). For quantum machine learning specifically, our Bures-preserving latent space enables geometry-aware kernel methods and variational classifiers that respect quantum statistical distance, potentially improving generalization when training data is limited by experimental constraints.

\section{Conclusion}
We introduced geometric latent space tomography, achieving high-fidelity state reconstruction and strong metric preservation between latent geodesics and the Bures metric. The learned 20-dimensional latent manifold exhibits three signatures of faithful geometric embedding: reduced intrinsic dimensionality, measurable Riemannian curvature, and strong correlation with quantum distances, preserving 78\% of Bures metric structure.

Our metric-preserving training objective demonstrates that geometric fidelity and tomographic accuracy are compatible, enabling geometry-aware operations like state discrimination, efficient fidelity estimation, and interpretable error manifolds without repeated full tomography (critically important for NISQ devices with limited coherence times).

Open questions include establishing rigorous approximation bounds, characterizing optimal trade-offs between latent dimension and metric preservation, and extending to many-qubit systems via classical shadows, tensor networks, or symmetry-equivariant architectures.

Our results suggest a broader paradigm: \textit{when data inhabit Riemannian manifolds with known geometric structure, neural representations should be constrained to preserve that structure}. By learning latent spaces respecting quantum geometry, we obtain representations that are simultaneously compact, accurate, and physically meaningful (essential tools as quantum computing advances toward practical advantage). Metric-preserving neural representations enable new forms of geometric quantum machine learning that leverage rather than ignore the profound structure of quantum state space.


\end{document}